\documentclass[prd,aps,a4paper,,twocolumn,eqsecnum,floatfix]{revtex4}  %,nofootinbib

\newif\ifusesec
\usesectrue  
   
\usepackage{graphicx} 
\usepackage{amsmath,amsfonts,amssymb}
\usepackage{mathtools}

\newcommand{\beq}{\begin{equation}}
\newcommand{\eeq}{\end{equation}}
\newcommand{\bea}{\begin{eqnarray}}
\newcommand{\eea}{\end{eqnarray}}
\begin{document}

\title{Radiation-reaction correction to   scattering binary dynamics at the Next-to-Leading Post-Newtonian Order }

\author{Donato Bini$^{1}$, Andrea Geralico$^{1}$, Sara Rufrano Aliberti$^{2,3}$}  
  \affiliation{
$^1$Istituto per le Applicazioni del Calcolo ``M. Picone'' CNR, I-00185 Rome, Italy\\
$^2$Scuola Superiore Meridionale, Largo San Marcellino 10, I-80138, Napoli, Italy\\
$^3$INFN, Sezione di Napoli, Complesso Universitario di Monte S. Angelo, Via Cinthia Edificio 6, I-80126, Napoli, Italy\\
}

\date{\today}

\begin{abstract}
We compute the next-to-leading-order radiation-reaction modification to the harmonic coordinate quasi-Keplerian parametrization of the binary dynamics, the two bodies undergoing a scattering process.
The solution for the radiation-reaction corrections to the orbital parameters is examined both in the time domain and in the frequency domain.
The knowledge of the radiation-reaction corrected orbit is a key ingredient for the calculation of the fractional 3.5PN corrections to the radiative losses as well as to the radiative multipole moments needed to build up the waveform at the same accuracy.
\end{abstract}

\maketitle

\section{Introduction}

Gravitational radiation-reaction effects play an important role in driving the dynamics of a binary system.
In a post-Newtonian (PN) framework radiation-reaction forces in the equations of motion arise at the 2.5PN order beyond the Newtonian acceleration. The complete dynamics at that leading order (LO) has been worked out in Refs. \cite{Damour:1981bh,DD1981a,D1982,Schaefer:1985vxb,kopei, Blanchet:1998vx}, where the full equations of motion are derived from a direct integration of the retarded field generated by the source.

Next-to-leading order (NLO) (i.e., 3.5PN) corrections in the radiation-reaction force have then been obtained by using a matching between the near-zone field and the wave-zone field \cite{Blanchet:1996vx,Jaranowski:1996nv,Pati:2002ux,Konigsdorffer:2003ue,Nissanke:2004er,Itoh:2009rz}. 
A different approach (restricted to the center-of-mass (CM) and leading to the same results) consists of requiring a balance between the (known) fluxes of energy and angular momentum radiated by gravitational waves at infinity and the losses of energy and angular momentum in the local equations of motion of the binary system \cite{Iyer:1993xi,Iyer:1995rn}. However, the balance equations do not completely fix the radiation-reaction force in the center-of-mass   frame at each order, and the residual coordinate freedom is encoded in a set of arbitrary gauge parameters.

The method of balance equations has been extended to the 4.5PN order in Ref. \cite{Gopakumar:1997ng} by assuming locality of the radiation-reaction force in the CM.
In contrast, at that order the radiation-reaction force acquires a non-local-in-time contribution which is responsible for the recoil of the system \footnote{The nonlocality of the 4PN tail term, however, differs from the one at 4.5PN in the CM frame: the latter indeed is \lq\lq semi-nonlocal," i.e., obtained as anti-derivatives of local terms.}, as pointed out in Ref. \cite{Blanchet:2024loi}. The explicit expression for the radiation-reaction force including nonlocal effects has then been obtained there in an extended Burke-Thorne gauge.
Notice that nonlocal effects in the radiation-reaction force first appear at the relative 1.5PN order (i.e., absolute 4PN) due to the presence of gravitational wave tails as non-local integrals, depending on the full past history of the system \cite{Blanchet:1987wq}.
Fractional 2PN order corrections to the radiation-reaction force have also been obtained within the effective-field-theory approach (including tails) \cite{Galley:2009px,Galley:2012qs,Galley:2015kus,Leibovich:2023xpg} as well as the Effective-One-Body (EOB) formalism \cite{Bini:2012ji}.

The effect of the presence of a radiation-reaction force in the equations of motion is the modification of the conservative dynamics. The latter can be parametrized in a quasi-Keplerian form \cite{dd,Damour:1988mr,SW1993,Memmesheimer:2004cv,Cho:2018upo} in the CM frame by introducing polar coordinates in the orbital plane and a set of orbital elements. At the 1PN level the parametrization looks Keplerian, with the orbital elements which in the bound case are given by the (inverse of the) radial period (or \lq\lq mean motion"), $n$, the periastron advance, $k$, the semi-major axis, $a_r$, and three eccentricities, $e_r$, $e_t$ and $e_\phi$, which differ from each other by PN corrections. 
The corresponding parametrization in the unbound case can then be obtained by analytic continuation \cite{dd}.
This is no more true starting at the 2PN order. Ref. \cite{Cho:2018upo} devised a version of the quasi-Keplerian representation which admits an analytic continuation at 2PN and 3PN levels.
In addition, the expressions of the orbital elements as functions of the conserved energy and angular momentum of the system depend on the chosen coordinates (typically harmonic or Arnowitt, Deser, and Misner (ADM)-type). 

A method to derive the radiation-reaction correction to the ellipticlike orbit of a binary system at the leading 2.5PN order was 
discussed in Ref. \cite{Damour:2004bz} based on the general Lagrange method of variation of arbitrary constants, later extended to the 3.5PN order in Ref. \cite{Konigsdorffer:2006zt}.
This method treats the radiation-reaction component of the binary relative acceleration as a first-order perturbation of the conservative piece.
The conservative problem admits in general four integrals of motion, which reduce to two when the problem is restricted to the orbital plane.
The unperturbed solution to the equations of motion in the CM frame can then be cast in a functional form which may be expressed in terms of two basic angles, so that it finally depends on four integration constants (the two integrals of motion plus the two initial values of the angles).
The problem turns out to be confined to a plane even in the presence of radiation reaction.
The solution of the perturbed system is thus obtained by keeping exactly the same functional form as in the unperturbed case, but allowing the four integration constants to be time-dependent, satisfying a set of first order evolution equations.

A hyperbolic version of this variation-of-constants method has been presented in Ref. \cite{Bini:2022enm}, where the 2.5PN correction to hyperbolic motion caused by the leading-order radiation-reaction force has been computed by using harmonic coordinates.
Starting from the 3.5PN radiation-force in harmonic coordinates derived in Ref. \cite{Nissanke:2004er}, we extend here the results of Ref. \cite{Bini:2022enm} to the NLO (i.e., 3.5PN order) by obtaining fractional 1PN corrections (in the CM frame).
We find that the solution for the perturbed orbital elements can be expressed in terms of elementary functions, whereas the correction to the orbit also involves polylogarithms.
We also compute the Fourier transform of the orbital elements, which in the soft limit gives the total variation during the whole scattering process.
The total change of the azimuthal angle gives instead the radiation-reaction contribution to the (relative) scattering angle, which agrees with known results \cite{Bini:2022enm}.

\subsection*{Notation}

Let us consider a nonspinning two-body system with masses $m_1$ and $m_2<m_1$ (total mass $M=m_1+m_2$, reduced mass $\mu=m_1m_2/M$, symmetric mass ratio $\nu=\mu/M$). 
Following the EOB formalism \cite{Buonanno:1998gg,Damour:2000we} we introduce the total energy of the system in the form
\beq
E_{\rm tot}\equiv E =Mc^2 h \,,\quad h=\sqrt{1+2\nu(\gamma-1)}\,,
\eeq
with $\gamma$ the corresponding effective energy (per unit of $\mu c^2$) and with binding energy
\beq
\bar E= \frac{h-1}{\nu}\,.
\eeq
We also use the  standard notation $p_\infty\equiv \sqrt{\gamma^2-1}$, and consider $p_\infty$ as a convenient energy variable to be used in PN-expanded quantities.

Concerning the parametrization of the orbit, we shall find it convenient to work with a dimensionless radial distance $r= c^2 r^{\rm phys}/(GM)$ and dimensionless rescaled orbital parameters, e.g., $\bar a_r \equiv c^2 \bar a^{\rm phys}/( GM)$ for the (hyperbolic motion analogue of the) semi-major axis. 

To ease notation we will generally set $G=1=c$, but use $\eta=\frac{1}{c}$ as a place-holder for PN expansions \footnote{Previous works on the same topic, e.g. Ref. \cite{Iyer:1993xi}, used the symbol $\eta$ to denote the symmetric mass ratio. The latter is nowadays universally denoted as $\nu$ and hence having $\eta$ as a place-holder of $1/c$ should not create confusion.}.

\section{Radiation-reaction correction to the two-body dynamics}

In general, the two-body's relative acceleration ${\mathbf a}={\mathbf a}_1-{\mathbf a}_2$ in the center-of-mass frame reads 
\bea
{\mathbf a}&=&{\mathbf a}_{\rm N}+\eta^2 {\mathbf a}_{\rm 1PN}+\eta^4 {\mathbf a}_{\rm 2PN}+\eta^5 {\mathbf A}^{\rm rr}_{\rm 2.5PN}\nonumber\\
&+&\eta^6 {\mathbf a}_{\rm 3PN}+\eta^7 {\mathbf A}^{\rm rr}_{\rm 3.5PN}
+\eta^8 {\mathbf a}_{\rm 4PN}^{\rm inst+tail}+\eta^9 {\mathbf A}^{\rm rr}_{\rm 4.5PN}\nonumber\\
&+&\eta^{10}  {\mathbf a}_{\rm 5PN}^{\rm inst+tail+rr}  
+\ldots\,,
\eea
where at 4PN the acceleration ${\mathbf a}_{\rm 4PN}^{\rm inst+tail}$ includes both instantaneous contributions and those due to tails (time-symmetric or conservative but also time-antisymmetric or dissipative \cite{Damour:2014jta});  similarly,  at 5PN one finds both the instantaneous conservative contribution and the radiation-reaction  ones.
It is customary to split the conservative and radiation-reaction parts of the acceleration
\beq
\label{asplit}
{\mathbf a}={\mathbf a}_{\rm cons} +{\mathbf A}_{\rm rr}\,,
\eeq
with
\bea
{\mathbf a}_{\rm cons}^i&=&  
\sum_k {\mathbf a}_{\rm kPN}^i \eta^{2k}+O(\eta^{11})\,,
\eea
with ${\mathbf a}_{\rm 0PN}={\mathbf a}_{\rm N}$, and
\bea
{\mathbf A}_{\rm rr}&=& \eta^5 {\mathbf A}^{\rm rr}_{\rm 2.5PN} +\eta^7 {\mathbf A}^{\rm rr}_{\rm 3.5PN}
 +\eta^9 {\mathbf A}^{\rm rr}_{\rm 4.5PN}\nonumber\\
&+&\eta^{10}  {\mathbf A}^{\rm rr}_{\rm 5PN} +O(\eta^{11})\,.
\eea 
Here, the $O(\eta^{11})$, i.e., 5.5PN terms contain both conservative and dissipative contributions.

\subsection{Conservative relative acceleration}

The conservative relative acceleration reads as
\bea
{\mathbf a}_{\rm nPN}&=& -\frac{M}{r^2}\left[{\mathcal A}_{\rm nPN}{\mathbf n}+{\mathcal B}_{\rm nPN}\dot r {\mathbf v}\right]\,;
\eea
for example, at the Newtonian level ${\mathbf a}_{\rm 0PN}={\mathbf a}_{\rm N}$ is given by
\beq
{\mathbf a}_{\rm N}=-\frac{M}{r^2}{\mathbf n}\,,
\eeq
that is ${\mathcal A}_{\rm N}=1$ and ${\mathcal B}_{\rm N}=0$, 
while at the 1PN level
\bea
{\mathcal A}_{\rm 1PN}&=&-2(2+\nu)\frac{M}{r}+(1+3\nu)v^2 -\frac32 \nu \dot r^2\,, \nonumber\\
{\mathcal B}_{\rm 1PN}&=&-2(2-\nu)\,.
\eea
At the 2PN level
\bea
{\mathcal A}_{\rm 2PN}&=&\frac34 (12+29\nu)\left(\frac{M}{r}\right)^2 +\nu(3-4\nu) v^4\nonumber\\ &+&\frac{15}{8}\nu(1-3\nu)\dot r^4
-\frac32 \nu (3-4\nu)v^2\dot r^2\nonumber\\
&-&\frac12 \nu (13-4\nu)\frac{M}{r}v^2 -(2+25\nu+2\nu^2)\frac{M}{r}\dot r^2 \,, \nonumber\\
{\mathcal B}_{\rm 2PN}&=&\frac12(4+41\nu+8\nu^2) \frac{M}{r} -\frac12\nu (15+4\nu)v^2\nonumber\\
&+&\frac32\nu(3+2\nu) \dot r^2\,.
\eea
The structure of ${\mathcal A}_{\rm nPN}$ and ${\mathcal B}_{\rm nPN}$  is similar when going to higher PN orders.

\subsection{Radiation-reaction relative acceleration}

Radiation-reaction (RR) effects arise at the 2.5 (absolute) PN level \cite{Iyer:1995rn}. The corresponding 2.5PN RR force in the CM frame has been computed within an arbitrary coordinate system, so that in general it depends on {\it two} gauge parameters, $\alpha$ and $\beta$, which can be fixed according to special coordinate choices. For example, in harmonic coordinates $(\alpha=-1, \beta=0)$; in ADM coordinates $(\alpha=\frac53, \beta=3)$; in Burke-Thorne (BT) coordinates $(\alpha=4, \beta=5)$ and in the EOB coordinates $(\alpha=0, \beta=2)$.

At the next 3.5PN level \cite{Nissanke:2004er}, the gauge parameters become {\it six} \cite{Iyer:1993xi}  and the expression of the RR force is still available in various coordinate systems, including the harmonic coordinate system. 
The gauge parameters corresponding to harmonic coordinates   are given by
\bea
\delta_1 &=& \frac{271}{28}+6\nu,\quad \delta_2 = -\frac{77}{4}-\frac32\nu,\quad \delta_3 = \frac{79}{14}-\frac{92}{7}\nu,\nonumber\\ 
\delta_4 &=& 10,\quad \delta_5 = \frac{5}{42}+\frac{242}{21}\nu,\quad \delta_6 = -\frac{439}{28}+\frac{18}{7}\nu,
\eea
following the standard notation of Ref. \cite{Nissanke:2004er}.
Proceeding beyond this level, i.e. at the 4.5PN level \cite{Blanchet:2024loi}, the RR force is explicitly known only in BT coordinates due to technical difficulties  appearing when performing the explicit computations in harmonic coordinates. The gauge parameter are instead twelve. The next level, 5PN, is currently an uncharted territory because one has to consider at this order  terms  quadratic   in the leading-order (LO) 2.5PN RR force.

In general, the CM radiation reaction relative force and acceleration, related as ${\mathbf F}^{\rm rr}=\mu {\mathbf A}^{\rm rr}$ write  as
\bea
\label{Arr}
{\mathbf A}^{\rm rr}&=&-\frac85 \nu \frac{G^2}{c^5} \frac{M^2}{r^3}\left[-(A_{2.5\rm PN}+\eta^2 A_{3.5\rm PN}+\ldots)\dot r {\mathbf n}\right.\nonumber\\
&&\left.+(B_{2.5\rm PN}+\eta^2 B_{3.5\rm PN}+\ldots){\mathbf v}\right]\nonumber\\
&=& {\mathbf A}^{\rm rr}_{\rm 2.5PN}+\eta^2 {\mathbf A}^{\rm rr}_{\rm 3.5PN}+\ldots\,.
\eea
As already said, ${\mathbf A}^{\rm rr}$ being a coordinate-dependent quantity, includes gauge parameters.
At the LO 2.5PN these parameters are denoted as $(\alpha, \beta)$ and we have
\bea
\label{AB_alpha_beta}
A_{2.5\rm PN}&=&3(1+\beta)v^2+\frac13 (23+6\alpha-9\beta)\frac{GM}{r}-5\beta \dot r^2\,, \nonumber\\
B_{2.5\rm PN}&=&(2+\alpha)v^2+(2-\alpha)\frac{GM}{r}-3(1+\alpha)\dot r^2 \,,\nonumber\\
\eea
In harmonic coordinates one has $\alpha=-1$ and $\beta=0$ (see Ref.  
\cite{dd}),
\bea
A_{2.5\rm PN,h}&=&3v^2+\frac{17}3 \frac{GM}{r}\,, \nonumber\\
B_{2.5\rm PN,h}&=&v^2+3\frac{GM}{r} \,.
\eea

At the 3.5PN level one writes
\bea
\label{AB_alpha_beta}
A_{3.5\rm PN}&=& c_1 v^4 +c_2 v^2 \frac{GM}{c^2r}+c_3 v^2 \dot r^2 +c_4 \dot r^2 \frac{GM}{r}\nonumber\\
&+&c_5 \dot r^4 +c_6 \left(\frac{GM}{r}\right)^2 \,,\nonumber\\
B_{3.5\rm PN}&=&  d_1 v^4 +d_2 v^2 \frac{GM}{r}+d_3 v^2 \dot r^2 +d_4 \dot r^2 \frac{GM}{r}\nonumber\\
&+& d_5 \dot r^4 +d_6 \left(\frac{GM}{r}\right)^2\,,
\eea
and each coefficient, corresponding to fractional 1PN corrections to the leading-order terms, has a linear-in-$\nu$ structure.
In this case, the coefficients $c_i$ and $d_i$, including gauge parameters, are listed in Table \ref{tab:table35rr}.

\begin{table*}  
\caption{\label{tab:table35rr}  Values of the coefficients $c_i$ and $d_i$ entering the radiation-reaction relative acceleration at the 3.5PN level of accuracy. The 3.5PN six gauge parameters are denoted as $\delta_1\ldots \delta_6$ following Ref. \cite{Iyer:1995rn}. $\alpha$ and $\beta$ are instead the 2.5PN two gauge parameters.}
\begin{ruledtabular}
\begin{tabular}{l|l}
$c_1$ & $\frac{117}{28} +  \frac{33}{7}\nu  -  \frac{3}{2}\beta(1 - 3\nu)  + 3\delta_2 - 3\delta_6$   \\  
$c_2$ & $-\frac{99}{14} +  \frac{155}{21}\nu  - 3\alpha(1 - 4\nu) -  \frac{3}{2}\beta (7 + 13\nu)  - 2\delta_1 - 3\delta_2 + 3\delta_5 + 3\delta_6$   \\  
$c_3$ & $\frac{95}{28} -  \frac{90}{7}\nu  +  \frac{5}{2}\beta (1 - 3\nu) - 5\delta_2 + 5\delta_4 + 5\delta_6$   \\  
$c_4$ & $-\frac{687}{28} +  \frac{92}{7}\nu   - 6\alpha\nu + \frac{\beta}{2} (54 + 17\nu) - 2\delta_2 - 5\delta_4 - 6\delta_5$   \\  
$c_5$ & $-7\delta_4$   \\  
$c_6$ & $-73 -  \frac{166}{7}\nu  - \alpha (14 + 9\nu) + 3\beta (7 + 4\nu) - 2\delta_3 - 3\delta_5$   \\  
\hline
$d_1$ & $ -3 + 9\nu -  \frac{3}{2}\alpha (1 - 3\nu)  - \delta_1$   \\  
$d_2$ & $-\frac{139}{84} -  \frac{64}{7}\nu  - \frac{\alpha}{2} (5 + 17\nu) + \delta_1 - \delta_3$   \\  
$d_3$ & $\frac{369}{28} -  \frac{156}{7}\nu  + \frac{3}{2} (3\alpha + 2\beta) (1 - 3\nu)  + 3\delta_1 - 3\delta_6$   \\  
$d_4$ & $\frac{295}{42} -  \frac{335}{42}\nu  + \frac{\alpha}{2} (38 - 11\nu)  - 3\beta (1 - 3\nu) + 2\delta_1 + 4\delta_3 + 3\delta_6$   \\  
$d_5$ & $\frac{95}{28} -  \frac{90}{7}\nu  - 5\beta (1 - 3\nu) + 5\delta_6$   \\  
$d_6$ & $-\frac{634}{21} +  \frac{22}{7}\nu  + \alpha (7 + 3\nu) + \delta_3$   \\  
\hline
\end{tabular}
\end{ruledtabular}
\end{table*}

The explicit expressions for the quantities \eqref{AB_alpha_beta} in harmonic coordinates then read
\begin{widetext}
\bea
A_{3.5\rm PN,h}&=&  \left(-\frac{183}{28}-\frac{15}{2}\nu \right)v^4+\left(-\frac{173}{14}+\frac{181}{6}\nu \right) v^2\frac{G M}{r}
+\left(\frac{285}{4}+\frac{15}{2}\nu \right) v^2 \dot r^2+\left(-\frac{147}{4}-47\nu \right) \dot r^2\frac{G M}{r}\nonumber\\
&-&70 \dot r^4+\left(-\frac{989}{14}-23\nu \right)\frac{G^2 M^2}{r^2}\,, \nonumber\\
B_{3.5\rm PN,h}&=&  \left(-\frac{313}{28}-\frac{3}{2}\nu \right) v^4+\left(\frac{205}{42}+\frac{37}{2}\nu \right) v^2\frac{G M}{r}
+\left(\frac{339}{4}+\frac{3}{2}\nu \right) v^2	\dot r^2+\left(-\frac{205}{12}-\frac{106}{3}\nu \right)\dot r^2\frac{G M}{r}-75\dot r^4\nonumber\\
&+&\left(-\frac{1325}{42}-13\nu\right)\frac{G^2 M^2}{r^2}  \,.
\eea
\end{widetext}

In the next section we will study modifications to the orbital parameters induced by the radiation reaction acceleration.

\section{Variation of constants method at the 3.5PN order}

Consider the gravitational two-body system as described in harmonic coordinates and in the center-of-mass frame under the presence of a radiation-reaction force.
In order to  study radiation-reaction effects, one can start from the conservative problem and use Lagrange's method of variation of constants to account for radiation-reaction effects, first applied in Ref. \cite{Damour:2004bz} to the case of bound motion, and later generalized to the unbound case in Ref. \cite{Bini:2022enm} as briefly recalled below.

The relative acceleration is split in an (unperturbed) integrable part (i.e., the conservative piece ${\mathbf a}_{\rm cons}$ in Eq. \eqref{asplit}) and a radiation-reaction part (${\mathbf A}_{\rm rr}$), which is treated as a perturbation.
The motion is confined to a plane even in the presence of radiation reaction, so that one can introduce polar coordinates.
The functional form for the {\it hyperboliclike} solution to the unperturbed (conservative) equations of motion reads
\bea
\label{sys0}
r&=& S(l,c_1,c_2)\,,\nonumber\\
\dot r&=&\bar n(c_1,c_2) \frac{\partial S(l,c_1,c_2)}{\partial l}
\,,\nonumber\\
\phi&=& c_\phi +{\overline W}(l,c_1,c_2) \,,\nonumber\\ 
\dot \phi&=&\bar n(c_1,c_2) \frac{\partial {\overline W}(l,c_1,c_2)}{\partial l}
\,,
\eea
where
\beq
l=\bar n(c_1,c_2)(t-t_0)+c_l\,,
\eeq
with $t_0$ denoting an arbitrary reference time value.
Therefore, the unperturbed solution depends on the four integration constants $c_1,c_2,c_l,c_\phi$.

The explicit expressions for the functions $S$ and $\bar W$ can be found by using the 3PN-accurate quasi-Keplerian parametrization of the hyperboliclike motion \cite{Cho:2018upo}
\begin{eqnarray} 
r&=& \bar a_r (e_r \cosh v-1)\,,\nonumber\\
\bar n\,  (t-t_0)&=&e_t \sinh v-v + f_t V+g_t \sin V\nonumber\\
&&
+h_t \sin 2V+i_t \sin 3V\,,\nonumber\\
\phi-\phi_0 &=&K[V+f_\phi \sin 2V+g_\phi \sin 3V\nonumber\\
&&
+h_\phi \sin 4V+i_\phi \sin 5V]\,,
\end{eqnarray}
with
\beq
\label{Vdef}
V(v)=2\, {\rm arctan}\left[\sqrt{\frac{e_\phi+1}{e_\phi-1}}\tanh \frac{v}{2}  \right]\,.
\eeq
The expressions for the orbital parameters in modified harmonic coordinates can be found in Appendix D of Ref. \cite{Bini:2022enm} as functions of the specific binding energy $\bar E \equiv (E-Mc^2)/(\mu c^2)$ and the dimensionless angular momentum $j=c J/(GM\mu)$ of the system, or equivalently in terms of $\bar a_r$ and $e_r$.
Equation \eqref{sys0} then implies 
\bea \label{SW}
S&=&\bar a_r (e_r \cosh v-1)
\,,\nonumber\\
{\overline W}&=&K[V+f_\phi \sin 2V+g_\phi \sin 3V\nonumber\\
&&
+h_\phi \sin 4V+i_\phi \sin 5V]\,, 
\eea
where the auxiliary variables $v$ must be expressed as a function of $l$, $c_1$ and $c_2$ by inverting the relation
\bea
l&=&e_t \sinh v-v + f_t V+g_t \sin V\nonumber\\
&&
+h_t \sin 2V+i_t \sin 3V\,.
\eea
In this paper we will only need the fractional 1PN accuracy on the hyperbolic motion, corresponding to 
\bea \label{SW}
S&=&\bar a_r (e_r \cosh v-1)
\,,\nonumber\\
{\overline W}&=&K V \,,\nonumber\\
l&=&e_t \sinh v-v \,.
\eea

The perturbed solution is then assumed to have the same functional form as in Eq. \eqref{sys0} (where the time dependence is entirely expressed via the variable $l$), but allowing the four constants $c_1$, $c_2$, $c_l$ and $c_\phi$ to be functions of time, i.e., $c_1(t)$, $c_2(t)$, $c_l(t)$ and $c_\phi(t)$ and the variable $l(t)$ to be given by
\beq
\label{ldef}
l(t)=\int_{t_0}^t\bar n(c_1(t),c_2(t)) dt+c_l(t)\,.
\eeq

The evolving constants $c_1(t)$, $c_2(t)$, $c_l(t)$ and $c_\phi(t)$ then satisfy the following set of equations
\bea \label{LagrangeHyp}
\frac{d c_1}{dt}&=&  \frac{\partial{c_1({\mathbf x},{\mathbf v})}}{\partial {\mathbf v}} \cdot {\mathbf A}_{\rm rr}
\,,\nonumber\\
\frac{d c_2}{dt}&=&  \frac{\partial{c_2({\mathbf x},{\mathbf v})}}{\partial {\mathbf v}} \cdot {\mathbf A}_{\rm rr}
\,,\nonumber\\
 \frac{d c_l}{dt} &=&-\left(\frac{\partial S}{\partial l}\right)^{-1}\left[\frac{\partial S}{\partial  c_1}\frac{d c_1}{dt}+\frac{\partial S}{\partial c_2}\frac{dc_2}{dt}\right]  
\,,\nonumber\\
\frac{d c_\phi}{dt} &=&-\frac{\partial {\overline W}}{\partial l}\frac{d c_l}{dt}-\frac{\partial {\overline W}}{\partial c_1}\frac{d  c_1}{dt}-\frac{\partial {\overline W}}{\partial c_2}\frac{d  c_2}{dt}\, ,
\eea
where ${\mathbf A}^{\rm rr}$ denotes the (relative) radiation-reaction acceleration \eqref{Arr}.

Notice that the quantities on the right hand side of Eqs. \eqref{LagrangeHyp} all take their unperturbed values, and for the purpose of the present paper need to be expanded up to the fractional 1PN order.

We choose $c_1=\bar a_r$ and $c_2=e_r$, which can be expressed in terms of the energy and angular momentum parameters $\bar E$ and $j$ as
\bea
\bar a_r&=&\frac{1}{2\bar E}\left[1-\frac{\bar E}{2}(-7+\nu)\eta^2+O(\eta^4)\right]
\,,\nonumber\\
e_r&=&\sqrt{1+2\bar Ej^2}\left[1+\frac{\bar E}{2}\frac{5\bar E j^2 (\nu - 3) + 2(\nu - 6)}{1+2\bar Ej^2}\eta^2\right.\nonumber\\
&&\left.
+O(\eta^4)\right]\,,
\eea
with
\bea
\bar E&=&  \frac12 v^2-\frac{G M}{r}
+\eta^2 \left[\frac38 (1-3\nu) v^4+\frac12 (3+\nu) v^2 \frac{G M}{r}\right.\nonumber\\
&&\left.
+\frac12 \nu\frac{G M}{ r} \dot r^2+\frac12 \frac{G^2 M^2}{r^2}\right] +O(\eta^4)
\,,\nonumber\\
j&=& (x v_y-y v_x) \left[1+\eta^2 \left(\frac12 (1-3\nu) v^2+(3+\nu)\frac{G M}{ r} \right)\right.\nonumber\\
&&\left.
 +O(\eta^4)\right]\,,
\eea
as functions of $x^i$ and $v^i$.  

When working at the leading 2.5PN absolute order, the evolution equations for $\bar a_r$ and $e_r$ are compact enough and read
\bea \label{Lagrange}
\frac{d \bar a_r}{dt}&=&-2\bar a_r^2{\mathbf v}\cdot {\mathbf A}_{\rm rr}^{\rm 2.5PN} 
\,,\nonumber\\
\frac{d e_r}{dt}&=&\frac{e_r^2-1}{e_r}\bar a_r{\mathbf v}\cdot {\mathbf A}_{\rm rr}^{\rm 2.5PN}
+\frac{\sqrt{e_r^2-1}}{e_r\sqrt{\bar a_r}} [{\mathbf x}\times {\mathbf A}_{\rm rr}^{\rm 2.5PN}]_z\,.\nonumber\\
\eea 
At the NLO, instead, when expressing all quantities as functions of position and velocities we decompose them as  
\bea
\bar a_r(x^i,v^i)&=&\bar a_r^{(0)}(x^i,v^i)+\eta^2 \bar a_r^{(2)}(x^i,v^i)\,,\nonumber\\
{\mathbf A}^{\rm rr}(x^i,v^i)&=& {\mathbf A}^{(0)\rm rr}(x^i,v^i)+\eta^2 {\mathbf A}^{(2)\rm rr}(x^i,v^i)\,,\qquad
\eea
with ${\mathbf A}^{(0)\rm rr}={\mathbf A}^{\rm rr}_{\rm 2.5PN}$ and ${\mathbf A}^{(2)\rm rr}={\mathbf A}^{\rm rr}_{\rm 3.5PN}$, and hence, for example
\bea
\frac{d\bar a_r^{(0)}}{dt}&=&\frac{\partial \bar a_r^{(0)}}{\partial v^i}{A}^{(0)\rm rr}_i\,,\nonumber\\
\frac{d\bar a_r^{(2)}}{dt}&=&\frac{\partial \bar a_r^{(0)}}{\partial v^i}{A}^{(2)\rm rr}_i +\frac{\partial \bar a_r^{(2)}}{\partial v^i}{A}^{(0)\rm rr}_i\,,
\eea
so that
\bea
\frac{d\bar a_r^{(0)}}{dt}&=&-2[\bar a_r^{(0)}]^2 v^i{A}^{(0)\rm rr}_i\,,\nonumber\\
\frac{d\bar a_r^{(2)}}{dt}&=&-2[\bar a_r^{(0)}]^2 v^i{A}^{(2)\rm rr}_i+\frac{\partial \bar a_r^{(2)}}{\partial v^i}{A}^{(0)\rm rr}_i\,.
\eea

Recall that one should consider the auxiliary parameter $v$ as a function of $l$, $c_1$ and $c_2$, i.e., $v=v(l,\bar a_r,e_r)$,
so that by differentiating both sides of the 1PN relation $l=e_t \sinh v-v$ (with $e_t$ replaced in terms of $(c_1,c_2)=(\bar a_r, e_r)$), we find
\bea
\frac{\partial v}{\partial l} &=&  \frac{1}{{\mathcal X}}+\frac12 \frac{ e_r (3\nu-8)\cosh(v)}{\bar a_r {\mathcal X}^2}\eta^2\,,\nonumber\\
\frac{\partial v}{\partial \bar a_r} &=&  -\frac{(3\nu-8)e_r}{2\bar a_r^2} \frac{\sinh(v)}{{\mathcal X}}\eta^2\,,\nonumber\\
\frac{\partial v}{\partial e_r} &=&   -\frac{\sinh(v)}{{\mathcal X}}-\frac{(3\nu-8)}{2\bar a_r} \frac{\sinh(v)}{{\mathcal X}^2}\eta^2 \,,
\eea
having introduced the notation
\beq
\label{X_def}
{\mathcal X}\equiv {\mathcal X}(v)= e_r \cosh v-1\,.
\eeq

We decompose the four varying constants $c_\alpha(t)$ as
\beq
c_\alpha(t)=c_\alpha^0+\delta^{\rm rr}c_\alpha(t)\,,
\eeq
with the boundary conditions that $\delta^{\rm rr}c_\alpha(t)$ vanishes in the limit $t\to-\infty$.
The system of equations \eqref{LagrangeHyp} then yields  an evolution system for the radiation-reaction corrections $\delta^{\rm rr}c_\alpha(t)$, whose right-hand-side is given (at our approximation) by explicit functions of time involving  the unperturbed values $c_\alpha^0$, henceforth simply denoted by $c_\alpha$.
It is convenient to integrate perturbed quantities with respect to the auxiliary variable $v$ as follows
\beq
\delta^{\rm rr} c_\alpha(v)=\int_{-\infty}^v   \frac{dc_\alpha}{dt} \frac{dt}{dv}dv\,.
\eeq
The total variation between $t=-\infty$ and $t=+\infty$ is then simply given by the value at $t=+\infty$, which will be denoted by $[\delta^{\rm rr} c_\alpha]$.
The solutions for the radiation-reaction corrections $\delta^{\rm rr}c_\alpha(t)$ are listed in Table \ref{tab:1} in the form $\delta^{\rm rr}c_\alpha=\delta^{\rm rr}c_\alpha^{(0)}+\eta^2 \delta^{\rm rr}c_\alpha^{(2)}$, whereas their total variations are given in Appendix \ref{jumps}.

% table 1
\begin{table*}  
\caption{\label{tab:1}   3.5PN accurate solutions for the radiation reaction corrections $\delta^{\rm rr}c_\alpha(t)=\delta^{\rm rr}c_\alpha^{(0)}+\eta^2 \delta^{\rm rr}c_\alpha^{(2)}$. Integration constants are chosen so that $\lim_{t\to -\infty}\delta^{\rm rr}c_\alpha(t)=0$. 
The solutions for $\delta^{\rm rr}\bar n(t)$ and $\delta^{\rm rr}K(t)$ are also shown for completeness.
}
\begin{ruledtabular}
\begin{tabular}{ll}
$\delta^{\rm rr}\bar a_r^{(0)}$ & See Eq. (3.12) of Ref. \cite{Bini:2022enm}\\ 
$\delta^{\rm rr}\bar a_r^{(2)}$ &$\frac{\nu}{\bar a_r^{5/2}}\left[
\left(-\frac{32 (e_r^2-1)^2}{ {\mathcal X}^7} 
+\frac{(\frac{56}{3}\nu+\frac{274}{15}) (e_r^2-1)}{ {\mathcal X}^6} +
\frac{(-\frac{24}{5}\nu+\frac{12}{5}) e_r^2-24\nu-\frac{9166}{75} }{ {\mathcal X}^5}
+\frac{(-\frac{256}{15}\nu-\frac{157}{2}) e_r^2+\frac{1223}{25}+\frac{104}{15}\nu }{ {\mathcal X}^4(e_r^2-1)} \right.\right.$\\
&$+ \frac{(-\frac{16}{5}\nu-\frac{272}{35}) e_r^4+(\frac{112}{45}\nu-\frac{28277}{1050}) e_r^2-\frac{18017}{525}-\frac{344}{15}\nu }{ {\mathcal X}^3 (e_r^2-1)^2} +\frac{-\frac{3823}{420} e_r^4+(-\frac{23161}{175}-\frac{1124}{45}\nu) e_r^2-\frac{16327}{525}-\frac{512}{15} \nu }{ {\mathcal X}^2 (e_r^2-1)^3}$\\
&$+\left. \frac{(-\frac{235733}{2100}-\frac{352}{45}\nu) e_r^4+(-\frac{191726}{525}-\frac{4796}{45}\nu) e_r^2-\frac{21107}{525}-\frac{944}{15}\nu }{ {\mathcal X} (e_r^2-1)^4} \right)e_r \sinh v +\frac{(-\frac{235733}{2100}-\frac{352}{45}\nu) e_r^4+(-\frac{191726}{525}-\frac{4796}{45}\nu) e_r^2-\frac{21107}{525}-\frac{944}{15}\nu}{ (e_r^2-1)^4}$\\
&$+ \left. \frac{-\frac{3823}{210} e_r^6+(-\frac{328}{5}\nu-471) e_r^4+(-\frac{3472}{15}\nu-\frac{18476}{35}) e_r^2-\frac{1912}{105}-\frac{288}{5}\nu}{ (e_r^2-1)^{9/2})}At(v)\right]$\\ 
\hline
$\delta^{\rm rr}e_r^{(0)}$ & See Eq. (3.12) of Ref. \cite{Bini:2022enm}\\ 
$\delta^{\rm rr}e_r^{(2)}$ &$ \frac{\nu}{\bar a_r^{7/2}}\left[
\left(\frac{16(e_r^2-1)^3}{{\mathcal X}^7}
+\frac{(-\frac{137}{15}-\frac{28}{3}\nu) (e_r^2-1)^2}{{\mathcal X}^6}
+\frac{\left((\frac{114}{5}+\frac{12}{5}\nu) e_r^2+\frac{2783}{75}+12\nu\right) (e_r^2-1)}{{\mathcal X}^5}\right.\right.$\\
&$+  \frac{(\frac{2333}{60}+\frac{68}{15}\nu) e_r^2-\frac{1807}{75}+\frac{8}{15}\nu}{{\mathcal X}^4}
+\frac{(\frac{94}{35}+\frac{28}{5}\nu ) e_r^4+\left(\frac{223361}{6300}-\frac{296}{45}\nu \right) e_r^2-\frac{5717}{1575}+\frac{64}{5}\nu}{ {\mathcal X}^3 (e_r^2-1)}$\\
&$\left.+ \frac{\frac{8737}{840} e_r^4+(\frac{1042}{45}\nu+\frac{500513}{6300}) e_r^2+\frac{32}{5}\nu-\frac{5627}{1575}}{ {\mathcal X}^2 (e_r^2-1)^2}
+ \frac{-\frac{144}{35} e_r^6+(\frac{488}{45}\nu+\frac{1456529}{12600}) e_r^4+(\frac{3214}{45}\nu+\frac{135823}{900})e_r^2
+\frac{32}{5}\nu-\frac{5627}{1575}}{ {\mathcal X} (e_r^2-1)^3}\right)\sinh(v)$\\
%%%%
&$+ \frac{-\frac{144}{35} e_r^6+(\frac{488}{45}\nu+\frac{1456529}{12600}) e_r^4+(\frac{3214}{45}\nu+\frac{135823}{900}) e_r^2+\frac{32}{5}\nu-\frac{5627}{1575}}{ e_r (e_r^2-1)^3}
+ \left.\frac{\left(\frac{5281}{420} e_r^4+(\frac{2585}{7}+68\nu) e_r^2+\frac{14258}{105}+\frac{328}{3}\nu\right)e_r}{ (e_r^2-1)^{7/2}} At(v)\right]$\\ 
\hline
$\delta^{\rm rr}c_l^{(0)}$ &See Eq. (3.12) of Ref. \cite{Bini:2022enm}\\ 
$\delta^{\rm rr}c_l^{(2)}$ &$\frac{\nu}{\bar a_r^{7/2}}\left[(-\frac{16(e_r^2-1)^4}{ {\mathcal X}^7 e_r^2 }
+\frac{(\frac{28}{3}\nu+\frac{377}{15})(e_r^2-1)^3}{ {\mathcal X}^6 e_r^2} 
+\frac{\left((\frac{94}{5}-\frac{12}{5}\nu) e_r^2-\frac{64}{3}\nu-\frac{1156}{25}\right) (e_r^2-1)^2}{ {\mathcal X}^5 e_r^2}+\frac{\left((-\frac{244}{5}\nu+\frac{74}{5}) e_r^2+\frac{172}{15}\nu+\frac{306}{5}\right) (e_r^2-1)}{ {\mathcal X}^4 e_r^2} \right.$\\
&$+\left.
\frac{(\frac{1076}{105}-\frac{28}{5}\nu) e_r^4+(\frac{26738}{315}+\frac{3884}{45}\nu) e_r^2-\frac{6446}{315}-\frac{184}{15}\nu}{ {\mathcal X}^3 e_r^2}
+\frac{(-\frac{226}{35}+\frac{396}{5}\nu) e_r^2-\frac{32}{5}\nu+\frac{2}{35}}{ {\mathcal X}^2 e_r^2}
+\frac{\frac{128}{5}\nu-\frac{1248}{35}}{{\mathcal X}}\right]$\\ 
\hline
$\delta^{\rm rr}c_\phi^{(0)}$ &See Eq. (3.12) of Ref. \cite{Bini:2022enm}\\ 
$\delta^{\rm rr}c_\phi^{(2)}$ &$\frac{\nu}{\bar a_r^{7/2}}\left\{
-\frac{48 (7 e_r^2 + 8)}{5(e_r^2-1)^{7/2}}At^2(v)
+ \left[ \frac{672 (e_r^2 + \frac{8}{7}){\rm arctan}(\sqrt{\frac{e_r+1}{e_r-1}}) }{5(e_r^2-1)^{7/2}} 
- \left(\frac{ 96 (e_r^2 + \frac{13}{2}) }{ 5(e_r^2-1)^3 {\mathcal X} } 
+ \frac{48}{ (e_r^2 - 1)^2 {\mathcal X}^2 }\right) e_r\sinh(v)
\right] At(v)\right.$\\
&$+\left[\frac{96 (e_r^2 + \frac{13}{2})}{5(e_r^2-1)^3} 
+ \left(\frac{96(e_r^2 + \frac{13}{2})}{5(e_r^2-1)^3 {\mathcal X}} 
+ \frac{48}{ (e_r^2-1)^2 {\mathcal X}^2}\right)e_r\sinh(v)\right]{\rm arctan}(\sqrt{\frac{e_r+1}{e_r-1}})$\\
&$ -\frac{24 (2 e_r^2 + 13)}{5(e_r^2-1)^{5/2} {\mathcal X}} 
+ \frac{2 (119 e_r^4\nu + 83 e_r^4 - 231 e_r^2\nu - 334 e_r^2 + 112\nu + 41)}{35(e_r^2-1)^{3/2} e_r^2 {\mathcal X}^2} 
+\frac{ 2 (252 e_r^4\nu - 318 e_r^4 + 1750 e_r^2\nu + 6817 e_r^2 - 1932\nu - 4399)}{315(e_r^2-1)^{1/2} e_r^2 {\mathcal X}^3}$\\
&$\left.- \frac{(e_r^2-1)^{1/2} (117 e_r^2\nu + 158 e_r^2 - 172\nu - 1128)}{15 e_r^2 {\mathcal X}^4} 
- \frac{2 (e_r^2-1)^{3/2} (90 e_r^2\nu - 825 e_r^2 + 800\nu + 1734)}{75 e_r^2 {\mathcal X}^5} 
+ \frac{(e_r^2-1)^{5/2} (140\nu + 377)}{15 e_r^2 {\mathcal X}^6}- \frac{16(e_r^2-1)^{7/2}}{e_r^2 {\mathcal X}^7} \right\}$\\ 
\hline
$\delta^{\rm rr}\bar n^{(0)}$ &$\frac{\nu}{\bar a_r^4}\left[ 
-\frac{2(37e_r^4+292 e_r^2+96)}{ 5 (e_r^2-1)^{7/2}} At(v)
-\frac{ 673 e_r^2+602}{ 15  (e_r^2-1)^3 }-e_r \sinh(v)\left( 
 \frac{ 673 e_r^2+602}{ 15 (e_r^2-1)^3 {\mathcal X} }
+\frac{ 111 e_r^2+314}{ 15(e_r^2-1)^2  {\mathcal X}^2 } 
+\frac{2 (36 e_r^2+49)}{ 15 (e_r^2-1) {\mathcal X}^3 } 
+\frac{14}{{\mathcal X}^4}\right) 
\right] $\\ 
$\delta^{\rm rr}\bar n^{(2)}$ &$ \frac{\nu}{\bar a_r^5}\left\{e_r\sinh(v)\left[ 
\frac{48  (e_r^2-1)^2}{ {\mathcal X}^7}
+\frac{(-28\nu-\frac{137}{5})  (e_r^2-1)}{ {\mathcal X}^6}
+\frac{(\frac{36}{5}\nu-\frac{18}{5}) e_r^2+36\nu+\frac{4583}{25} }{{\mathcal X}^5}
+\frac{(\frac{51}{4}+\frac{559}{15}\nu) e_r^2-\frac{331}{15}\nu+\frac{1581}{50} }{ {\mathcal X}^4 (e_r^2-1)}
\right.\right.$\\
&$
+\frac{(-\frac{852}{35}+\frac{44}{5}\nu) e_r^4+(-\frac{103}{45}\nu+\frac{19177}{700}) e_r^2+\frac{1303}{45}\nu+\frac{35167}{350} }{{\mathcal X}^3 (e_r^2-1)^2}
+ \frac{(\frac{37}{6}\nu-\frac{11717}{280}) e_r^4+(\frac{16979}{175}+\frac{4387}{90}\nu) e_r^2+\frac{1519}{45}\nu+\frac{71277}{350} }{{\mathcal X}^2 (e_r^2-1)^3}$\\
&$+\left.\frac{(-\frac{235367}{1400}+\frac{4421}{90}\nu) e_r^4+(\frac{14033}{90}\nu+\frac{204151}{350}) e_r^2+\frac{126457}{350}+\frac{2743}{45}\nu }{{\mathcal X}(e_r^2-1)^4}\right]
+ \frac{(-\frac{235367}{1400}+\frac{4421}{90}\nu) e_r^4+(\frac{14033}{90}\nu+\frac{204151}{350}) e_r^2+\frac{126457}{350}+\frac{2743}{45}\nu}{ (e_r^2-1)^4}$\\
&$+\left. \frac{(-\frac{11717}{140}+\frac{37}{3}\nu)e_r^6+(\frac{917}{5}\nu-\frac{117}{2}) e_r^4+(\frac{4228}{15}\nu+\frac{48294}{35})e_r^2+\frac{272}{5}\nu+\frac{11036}{35}}{ (e_r^2-1)^{9/2}} At(v)
\right\}$\\
\hline
$\delta^{\rm rr}K^{(0)}$ &$ 0$\\ 
$\delta^{\rm rr}K^{(2)}$ &$ \frac{\nu}{\bar a_r^{7/2}}\left[
\frac{48 (7 e_r^2+8)}{ 5 (e_r^2-1)^{7/2}} At(v) 
+\frac{24 (2 e_r^2+13)}{ 5 (e_r^2-1)^3 }
+\left(\frac{24 }{ (e_r^2-1)^2 {\mathcal X}^2}
+\frac{24(2 e_r^2+13)}{ 5 {\mathcal X} (e_r^2-1)^3}\right)e_r\sinh(v) \right]$\\
\end{tabular}
\end{ruledtabular}
\end{table*}

Finally, the correction to the orbit is $r(t)=r^{\rm cons}(t)+\delta^{\rm rr} r(t)$, $\phi(t)=\phi^{\rm cons}(t)+\delta^{\rm rr} \phi(t)$ with
\bea
\delta^{\rm rr} r(t) &=& DS(l, \bar a_r, e_r)\,,\nonumber\\
\delta^{\rm rr} \phi(t) &=& \delta^{\rm rr} c_\phi(t) +D \bar W(l, \bar a_r, e_r)\,,
\eea
where we have introduced the notation 
\beq
\label{DX_notation}
D X(l, \bar a_r, e_r)= \frac{\partial X}{\partial l} \delta^{\rm rr} l(t)+\frac{\partial X}{\partial \bar a_r} \delta^{\rm rr} \bar a_r (t)
+\frac{\partial X}{\partial e_r} \delta^{\rm rr} e_r(t)\,,
\eeq
with
\bea
\label{deltarrldef}
 \delta^{\rm rr} l(t) &=& \int_{t_0}^t \delta^{\rm rr}\bar n(t) dt+ \delta^{\rm rr} c_l(t)\nonumber\\
&\equiv&\delta^{\rm rr} \tilde l(t)+\delta^{\rm rr} c_l(t)\,.
\eea

For instance, if $X={\overline W}=KV$ (at 1PN level) one has 
\begin{widetext}
\bea
\frac{\partial \bar W}{\partial l}&=&\frac{\sqrt{e_r^2-1}}{{\mathcal X}^2}
+ \left[\frac{2\sqrt{e_r^2-1} (\nu-2)}{{\mathcal X}^3}
+\frac{(14+3 e_r^2\nu-8 e_r^2-4\nu)}{2{\mathcal X}^2 \sqrt{e_r^2-1}} \right]\frac{\eta^2}{\bar a_r} 
\,,\nonumber\\
\frac{\partial \bar W}{\partial \bar a_r}&=& \left[-\frac{6}{e_r^2-1}{\rm arctan}\left(\alpha \tanh \frac{v}{2} \right)
-\frac{\nu e_r \sinh v}{2\sqrt{e_r^2-1}{\mathcal X}}
-\frac{e_r \sinh v \sqrt{e_r^2-1}(3\nu-8)}{2{\mathcal X}^2} \right]\frac{\eta^2}{\bar a_r^2}
\,,\nonumber\\
\frac{\partial \bar W}{\partial e_r} &=& -\frac{\sinh v}{\sqrt{e_r^2-1}{\mathcal X}}-\frac{\sinh v \sqrt{e_r^2-1}}{{\mathcal X}^2} \nonumber\\
&+& \left[-\frac{12e_r}{(e_r^2-1)^2} {\rm arctan}\left(\alpha \tanh \frac{v}{2} \right)
-\frac{2\sinh v \sqrt{e_r^2-1} (-2+\nu)}{{\mathcal X}^3}
-\frac{\sinh v (e_r^2\nu+6)}{2 (e_r^2-1)^{3/2}{\mathcal X}}  
-\frac{3\sinh v }{ \sqrt{e_r^2-1}{\mathcal X}^2} \right]\frac{\eta^2}{\bar a_r}
\,,\nonumber\\
\eea
with
\beq
\alpha=\sqrt{\frac{e_r+1}{e_r-1}}\,.
\eeq

The LO and NLO pieces of the explicit solution for $\delta^{\rm rr} l(t)$ are given by Eq. \eqref{deltarrldef} with $\delta^{\rm rr} \tilde l(t)=\delta^{\rm rr} \tilde l^{(0)}+\eta^2\delta^{\rm rr} \tilde l^{(2)}$, and 
\bea
\delta^{\rm rr} \tilde l^{(0)}&=&\frac{\nu}{\bar a_r^{5/2}}\left\{
\left[\ln({\mathcal X})-{\mathcal X}-1-e_r\sinh(v)+v+\ln(2)-\ln(e_r)\right]B
+\frac{2(49+36e_r^2)}{15{\mathcal X}(e_r^2-1)}+\frac{7}{{\mathcal X}^2}\right.\nonumber\\
&&\left.
+\left[-e_r\sinh(v)At(v)+\frac{i}{2}\left({\rm Li}_2(z)-{\rm Li}_2(\bar z)\right)-\frac{\sqrt{e_r^2-1}}{2}\right]C
\right\}
\,,
\eea
where (using the identity ${\rm arctan}(\sqrt{e_r^2-1})=-2{\rm arctan}(\alpha)+\pi$)
\beq
z\equiv \frac{e^v\, e_r}{ (1+i\sqrt{e_r^2-1})}=e^{v-i{\rm arctan}(\sqrt{e_r^2-1})}=-e^{v+2i{\rm arctan}(\alpha)},
\eeq
 and
\bea
B&=&\frac{602+673e_r^2}{15(e_r^2-1)^3}
\,,\nonumber\\
C&=&\frac{2(96+292e_r^2+37e_r^4)}{5(e_r^2-1)^{7/2}}
\,,
\eea
and
\bea
\delta^{\rm rr} \tilde l^{(2)}&=&\frac{\nu}{\bar a_r^{7/2}}\left\{
-\frac{i}{2}\left({\rm Li}_2(z)-{\rm Li}_2(\bar z)\right)D
+\left[e_r\sinh(v)At(v)+\frac{\sqrt{e_r^2-1}}{2}\right]E
+\left({\mathcal X}+1+e_r\sinh(v)\right)F
\right.\nonumber\\
&&
-\left(v+\ln({\mathcal X})+\ln(2)-\ln(e_r)\right)G
-\frac{e_r^4 \left(\frac{68 \nu
   }{5}-\frac{768}{35}\right)+e_r^2 \left(\frac{949 \nu
   }{90}-\frac{2809}{2100}\right)+\frac{2422 \nu
   }{45}+\frac{4717}{350}}{(e_r^2-1)^2{\mathcal X}}\nonumber\\
&&\left.
-\frac{e_r^2 \left(\frac{877 \nu
   }{30}+\frac{11}{40}\right)-\frac{197 \nu }{15}-\frac{227}{300}}{(e_r^2-1){\mathcal X}^2}
-\frac{e_r^2 \left(\frac{12 \nu }{5}-\frac{6}{5}\right)+19 \nu
   +\frac{1061}{25}}{{\mathcal X}^3}
+\frac{(e_r^2-1)(137+140\nu)}{20{\mathcal X}^4}
-\frac{48(e_r^2-1)^2}{5{\mathcal X}^5}
\right\}
\,,\nonumber\\
\eea
with
\bea
D-E&=&-\frac12(3\nu-8)C
\,,\nonumber\\
F-G&=&\frac12(3\nu-8)B
\,,\nonumber\\
E&=&\frac{e_r^6 \left(\frac{407 \nu
   }{15}-\frac{10681}{140}\right)+e_r^4 \left(\frac{1427
   \nu }{5}-\frac{15}{2}\right)+e_r^2 \left(\frac{3052 \nu
   }{15}+\frac{46922}{35}\right)+16 \nu +\frac{10364}{35}}{(e_r^2-1)^{9/2}}
\,,\nonumber\\
G&=&\frac{e_r^4 \left(\frac{1201 \nu
   }{45}+\frac{47293}{1400}\right)+e_r^2 \left(\frac{7123
   \nu }{45}+\frac{98348}{175}\right)+\frac{3646 \nu
   }{45}+\frac{63247}{350}}{(e_r^2-1)^4}\,.
\eea
\end{widetext}

Here we have used the notation
\beq
{\rm At}(v)\equiv {\rm arctan}\left(\alpha \tanh \frac{v}{2} \right)+{\rm arctan}(\alpha)\,.
\eeq
Notice also that the quantity 
\beq
-\frac{i}{2}\left({\rm Li}_2(z)-{\rm Li}_2(\bar z)\right)
\eeq
is real valued.
In fact, by using the properties of the dilogarithm of complex argument \cite{lewin} it can be written in terms of the Clausen function of the second order as follows
\bea
&&-\frac{i}{2}\left({\rm Li}_2(z)-{\rm Li}_2(\bar z)\right)=
\xi\ln r+\frac12{\rm Cl}_2(2\xi)+\frac12{\rm Cl}_2(2\theta)\nonumber\\
&&
-\frac12{\rm Cl}_2(2\xi+2\theta)\,,
\eea
where $z=re^{i\theta}$, with $r=e^v$ and $\theta=-{\rm arctan}(\sqrt{e_r^2-1})$, and $\tan\xi=r\sin\theta/(1-r\cos\theta)$, with $[\cos\theta,\sin\theta]=[1/e_r,-\sqrt{e_r^2-1}/e_r]$.
The beginning of its large-eccentricity expansion reads
\bea
&&-\frac{i}{2}\left({\rm Li}_2(z)-{\rm Li}_2(\bar z)\right)=
-{\rm Ti}_2(e^v)-\frac{v+\ln(2\cosh v)}{2e_r}\nonumber\\
&&+\frac{1}{4\cosh v\, e_r^2}
-\frac{1}{12e_r^3}\left(v+\ln(2\cosh v)-\frac{1}{\cosh^2v}\right)\nonumber\\
&&
+O\left(\frac{1}{e_r^4}\right)\,,
\eea
where
\beq
{\rm Ti}_2(x)=-\frac{i}{2}\left({\rm Li}_2(ix)-{\rm Li}_2(-ix)\right)\,,
\eeq
is the inverse tangent integral.

The total change of the azimuthal angle $[\delta^{\rm rr}\phi]$ gives the radiation-reaction contribution to the (relative) scattering angle,
\beq
\chi_{\rm rr}^{\rm 2.5PN+3.5PN}=[\delta^{\rm rr} \phi]^{(0)}+\eta^2[\delta^{\rm rr} \phi]^{(2)}\,,
\eeq
where
\begin{widetext}
\bea
{}[\delta^{\rm rr} \phi]^{(0)}&=& \frac{\nu}{\bar a_r^{5/2}}\left[\frac{2(121 e_r^2+304)}{ 15 (e_r^2-1)^3}  {\rm arccos}\left(-\frac{1}{e_r}\right)  
+\frac{2 (72 e_r^4+1069 e_r^2+134)}{ 45e_r^2 (e_r^2-1)^{5/2}}  \right]
\,,\nonumber\\
{}[\delta^{\rm rr} \phi]^{(2)}&=& \frac{\nu}{\bar a_r^{7/2}}\left[ \frac{48(7 e_r^2+8)}{ 5(e_r^2-1)^{7/2}}  {\rm arccos}^2\left(-\frac{1}{e_r}\right)
\right.\nonumber\\
&&+\frac{(1316 \nu+2783) e_r^4-( 36400  \nu+90420) e_r^2-51296\nu-58376}{ 420(e_r^2-1)^4}  {\rm arccos}\left(-\frac{1}{e_r}\right) \nonumber\\
&&+\left. \frac{(10080 \nu+51840) e_r^6-(175420\nu+1396049) e_r^4-(1049720 \nu +1003562)e_r^2-80640\nu+157576}{ 6300 e_r^2 (e_r^2-1)^{7/2}}\right]\,,\nonumber\\
\eea
which agrees with previous results (see Eqs. (C8)--(C9) of Ref. \cite{Bini:2022enm}).
The corresponding expression in terms of $(p_\infty, j)$ in a large-$j$ expansion limit is then
\bea
\chi_{\rm rr}^{\rm 2.5PN+3.5PN}&=&\nu\left\{\left[
\left(\frac{80}{7}  -\frac{16}{5}\nu \right)\frac{p_\infty^4}{j^3}
+\left(\frac{23111}{840} -\frac{437}{30}\nu \right)\frac{p_\infty^3\pi}{j^4}
+\left(\frac{84}{5} \pi^2+\frac{86368}{525}  -\frac{8704}{45}\nu \right)\frac{p_\infty^2}{j^5}\right.\right.\nonumber\\
&+&\left.\left(\frac{15679}{60} -\frac{424}{3}\nu \right)\frac{p_\infty \pi}{ j^6}
+\left(36 \pi^2-\frac{5776}{9}\nu +\frac{1587856}{1575} \right)\frac{1}{j^7}\right]\eta^2\nonumber\\
&+& \left.\frac{16}{5} \frac{p_\infty^2}{j^3}
+\frac{121}{15} \frac{p_\infty \pi}{j^4}
+\frac{3008}{45}  \frac{1}{j^5}
+\frac{85}{3} \frac{\pi}{ p_\infty j^6}
+\frac{2576}{45} \frac{1}{p_\infty^2 j^7}\right\} +O\left(\frac{1}{j^8}\right)\,.
\eea
\end{widetext}

\section{Fourier space analysis of radiation reacted expressions}

We have expressed all radiation reacted orbital parameters in terms of $\delta^{\rm rr}\bar a_r$ and $\delta^{\rm rr} e_r$, to be considered as building elements.
In this section we will explore the Fourier space  representation of $\delta^{\rm rr}\bar a_r$ and $\delta^{\rm rr} e_r$ as well as of the other orbital elements, namely in a compact form $f(t)=[\delta^{\rm rr}\bar a_r,\delta^{\rm rr}e_r,\delta^{\rm rr} K, \delta^{\rm rr}\bar n]$. We will use the following notation for the Fourier transform (FT) of a function $f(t)$
\bea
\label{FT}
\hat f(\omega)&=&\int_{-\infty}^\infty dt e^{i\omega t} f(t)\,,
\eea
so that, for example,
\bea
\hat \delta^{\rm rr}\bar a_r(\omega)&=&\int_{-\infty}^\infty dt e^{i\omega t} \delta^{\rm rr}\bar a_r(t)\,.
\eea
The radiation-reaction corrections to the orbital elements are expressed in terms of the auxiliary variable $v$, so that Eq. \eqref{FT} can be rewritten as
\bea
\hat f(\omega)&=&\int_{-\infty}^\infty dv \frac{dt}{dv} e^{i\omega t(v)} f(t(v))\,.
\eea
Notice that as $t \to +\infty$ (i.e., $v\to +\infty$) one has $f(t(v))$ tends to the total variation of $f$, denoted as  $[f]$, during the whole scattering process (see Appendix A):
\beq
\label{totvar}
[f ]=\lim_{t\to \infty}f(v(t))=\int_{-\infty}^\infty \frac{df}{dt}\frac{dt}{dv} dv\,.
\eeq
As a consequence, the soft limit $\omega \to 0$ of $\hat f (\omega)$ should be of the type
\beq
\hat f (\omega) \overset{\omega \to 0}{=} \frac{i [f]}{\omega}\,.
\eeq

All the FT integrals are performed by taking a large-$e_r$ (or equivalently a large-$j$) expansion of the integrand, corresponding to a PM expansion.
Doing so it proves convenient to use the integration variable
\beq
T=\sinh v\,,\qquad dv=\frac{dT}{\sqrt{1+T^2}}\,,
\eeq
so that
\bea
\hat f (\omega)&=&\int_{-\infty}^\infty  \frac{dT}{\sqrt{1+T^2}} \frac{dt}{dv} e^{i\omega t(v)} f(t(v))\bigg|_{v={\rm arcsinh} T}\,,
\eea
and to replace $\omega$ by the new frequency variable
\beq
u\equiv \frac{\omega j}{p_\infty^2}\,,
\eeq
assuming $u>0$ for simplicity.
In particular, this condition removes eventual $\delta(u)$ terms in the FT which correspond to constant contributions in the time domain. 

We show in Table \ref{tab:3} the leading PN approximation of the first three  terms of the large-$j$ expansion of the FT of the radiation-reacted orbital elements.  The  Supplemental Material gives the full fractionally 1PN accurate expression of the first three terms of the large-$j$ expansion \cite{sup_mat}.

Our final results involve the following master integrals: $I_n$, $I_n^{\rm at}(u)$, $I_n^{\rm as}(u)$, $I_n^{\rm asat}(u)$.
 
The families of integrals
\bea
I_{n}(u)&=& \int_{-\infty}^\infty dT \frac{e^{iuT}}{(1+T^2)^{n/2}}\nonumber\\
&=&\sqrt{\pi} \frac{ 2^{\frac{3-n}{2}} u^{\frac{n-1}{2}}
K_{\frac{n-1}{2}}(u)   
 }{\Gamma[\frac{n}{2}]}\,,
\eea
and
\beq
I_n^{\rm at}(u)= \int_{-\infty}^\infty dT \frac{e^{iuT}{\rm arctan(T)}}{(1+T^2)^{n/2}}\,,
\eeq
have been already discussed in Ref. \cite{Bini:2024ijq}.
In particular, the integrals $I_n^{\rm at}(u)$ can be expressed in terms of \lq\lq iterated Bessel functions."
[See Appendix A there for the recursive evaluation of these integrals and their explicit expressions in terms of special functions.]
For example,
\bea
I_1(u) &=& 2K_0(u)  \,,\nonumber\\
I_2(u) &=& \pi e^{-u}\,,\nonumber\\
I_3(u) &=& 2 u K_1(u)\,,
\eea
etc., while
\bea
I_0^{\rm at}(u)&=& i\pi \frac{e^{-u}}{u}=\frac{i}{u}I_2(u) \,,
\eea
and
\bea
I_1^{\rm at}(u)&=&i\left[-\frac{\sqrt{\pi }}{2} I_0(u) G_{2,4}^{3,1}\left(u^2|
\begin{array}{c}
 1,1 \\
 \frac{1}{2},\frac{1}{2},\frac{1}{2},0 \\
\end{array}
\right)\right.\nonumber\\
&+&\frac{K_0(u) }{2 \sqrt{\pi }}G_{2,4}^{2,2}\left(u^2|
\begin{array}{c}
 1,1 \\
 \frac{1}{2},\frac{1}{2},0,\frac{1}{2} \\
\end{array}
\right)\nonumber\\
&+&\left. \frac{\pi ^2 I_0(u)}{2}\right]\,,
\eea
where $G$ denotes the Meijer G function.
 
Our results also involve integrals of the type
\bea
\label{Inasat}
I_n^{\rm as}(u)&=& \int_{-\infty}^\infty dT \frac{e^{iuT}{\rm arcsinh(T)}}{(1+T^2)^{n/2}}\,,\nonumber\\
I_n^{\rm asat}(u)&=& \int_{-\infty}^\infty dT \frac{e^{iuT}{\rm arcsinh(T)}{\rm arctan(T)}}{(1+T^2)^{n/2}}\,,
\eea
(similar integrals, involving higher powers of arcsinh and arctan occur as soon as the PN accuracy increases \cite{Bini:2024ijq}). Only some of the integrals Eq. \eqref{Inasat} can be explicitly evaluated in terms of known functions. E.g.
\bea
I_0^{\rm as}(u)&=& \frac{2i}{u} K_0(u)\,,\nonumber\\
I_1^{\rm as}(u)&=& i\pi K_0(u)\,,\nonumber\\
I_3^{\rm as}(u)&=& -i\pi [e^{-u}-u K_1(u)]\,.
\eea

The integral $I_0^{\rm asat}(u)$ can be performed explicitly too, starting from the identity
\beq
0=\int_{-\infty}^\infty \frac{d}{dT}\left[e^{iuT} {\rm arctan}(T){\rm arcsinh}(T)  \right]\,,
\eeq
which expanding the derivatives gives
\beq
I_0^{\rm asat}(u)=\frac{i}{u}(I_2^{\rm as}(u)+I_1^{\rm at}(u))\,.
\eeq
When going beyond the present 1PN (fractional) approximation and also expanding in large $j$ the radiation reacted expressions for the orbital elements will involve other special integrals (not needed here however) which are of the type
\beq
I_{m,n,p}(u)=\int dT e^{iuT} \frac{{\rm arcsinh}^m(T){\rm arctan}^n(T)}{(1+T^2)^p}\,.
\eeq
In order to evaluate them it is convenient to use recurrence relations. For example, the identity
\beq
0=\int dT \frac{d}{dT} \left[e^{iuT} \frac{{\rm arcsinh}^m(T){\rm arctan}^n(T)}{(1+T^2)^p}\right]
\eeq
implies the recursive relation
\bea
0&=&iu I_{m,n,p}(u)+n I_{m,n-1,p+1}(u)+m I_{m-1,n,p+\frac12}(u)\nonumber\\
&&
+2ip \frac{d}{du}I_{m,n,p+1}(u)\,.
\eea

Note also that integrals of the type $I_1^{\rm as, n}(u)$,
\beq
I_1^{\rm as, n}(u)=\int_{-\infty}^\infty dT \frac{e^{iuT}{\rm arcsinh}^n(T)}{(1+T^2)^{1/2}}
\eeq
are simply related to the derivatives with respect to the order of the Hankel functions
\beq
i\pi H_p^{(1)}(q)=\int_{-\infty}^\infty d\xi\, e^{q\sinh \xi-p\xi}\,,
\eeq
whose relation with the Bessel K functions is
\beq
H_p^{(1)}(iu)=\frac{2}{\pi} e^{-i(p+1)\frac{\pi}{2}}K_p(u)\,.
\eeq
In fact, changing the variable as $T=\sinh \xi$ one finds
\begin{eqnarray}
i\pi H_p^{(1)}(q)&=&\int_{-\infty}^\infty \frac{dT}{\sqrt{1+T^2}}\, e^{qT-p\, {\rm arcsinh}(T)}\nonumber\\
&=&\sum_{k=0}^\infty \frac{(-1)^k p^k }{k!} \int_{-\infty}^\infty \frac{dT}{\sqrt{1+T^2}}\, e^{qT}  {\rm arcsinh}^k(T)\nonumber\\
&=&\sum_{k=0}^\infty \frac{(-1)^k p^k }{k!} I_1^{\rm as, k} \,,
\end{eqnarray}
so that different powers of arcsinh (at the right-hand-side) correspond to higher-order derivatives of the Hankel functions with respect to the order (at the left-hand-side), unknown beyond the third order and much involved.

% table 3
\begin{table*}  
\caption{\label{tab:3} Fourier transform of radiation reacted orbital elements expressed in terms of the rescaled frequency $u=\frac{\omega p_\infty}{j^2}$ (assumed positive) and represented is a large-$j$ series expansion limit, explicitly showing the first three terms in the leading PN approximation. The complete 1PN expressions are included in the Supplemental Material \cite{sup_mat}.
}
\begin{ruledtabular}
\begin{tabular}{l|l}
$\hat  \delta^{rr}  K(u) \Big|_{\rm 3.5PN}$ &$  \nu\eta^2 \left[
\frac{96 ip_\infty K_1(u)}{ 5j^3} 
+\frac{(\frac{168}{5 u}+24)i\pi e^{-u}-\frac{96}{5}K_0(u)}{j^4}
+\frac{(-\frac{168}{5}\pi +48i u)K_0(u) +\frac{1104}{5}i K_1(u)-\frac{168}{ 5}I_1^{\rm at}(u)}{p_\infty j^5}
+O\left(\frac1{j^6}\right)\right]$\\
\hline
$\hat  \delta^{rr} e_r(u) \Big|_{\rm 2.5PN} $ &$\nu\left[ (-\frac{16}{5} K_1(u)-\frac{16}{5} u K_0(u))i p_\infty^2 
+[\frac{16}{5} K_0(u)+(-\frac{121}{ 15 u}-\frac{59}{15 } u -\frac{97}{15} )i \pi e^{-u}]\frac{p_\infty}{j} \right.$\\
&$\left.+\left((-\frac{778}{45} i u+\frac{121}{15}\pi)K_0(u)+\frac{121}{15} I_1^{\rm at}(u)+(-\frac{3152}{45} -\frac{188}{15} u^2)i K_1(u)\right)\frac{1}{j^2}
+O\left(\frac1{j^3}\right)\right]$\\
\hline
$\hat  \delta^{rr} \bar a_r(u) \Big|_{\rm 2.5PN} $ &$\nu \left[
\frac{32}{5}  u K_0(u)\frac{i}{p_\infty j}
+\left(\frac{74}{ 15 u} +\frac{118}{15} u +\frac{74}{15}  \right) \frac{i \pi e^{-u}}{p_\infty^2 j^2}\right. $\\
&$\left.+\left((\frac{196}{9} i u-\frac{74}{15}\pi)K_0(u)+(\frac{376}{15}i u^2+\frac{3136}{45} i)K_1(u)-\frac{74}{15}  I_1^{\rm at}(u)\right)\frac{1}{ p_\infty^3 j^3}
+O\left(\frac1{j^4}\right)\right]$\\
\hline
$\hat  \delta^{rr} \bar n (u) \Big|_{\rm 2.5PN} $ &$\nu\left[-\frac{48}{5}i  uK_0(u)\frac{p_\infty^4}{j}
+\left(-\frac{37}{5} -\frac{59}{5} u -\frac{37}{ 5 u} \right)\frac{i\pi e^{-u}p_\infty^3}{j^2} \right. $\\
&$\left. +\left((-\frac{98}{3} iu+\frac{37}{5}\pi)K_0(u)+(-\frac{188}{5}iu^2-\frac{1568}{15}i)K_1(u)+\frac{37}{5} I_1^{\rm at}(u)\right)\frac{ p_\infty^2}{ j^3}
+O\left(\frac1{j^4}\right)\right]$\\
\end{tabular}
\end{ruledtabular}
\end{table*}

\section{Concluding remarks}

We have used the variation of constants method introduced in Ref. \cite{Damour:2004bz} to compute the radiation-reaction modification to the harmonic coordinate quasi-Keplerian parametrization of the hyperboliclike two-body dynamics at the NLO, i.e., at the 3.5PN level, generalizing previous results at the leading (2.5PN) order \cite{Bini:2022enm}.
We have obtained the explicit time dependence of the radiation-reaction corrections to the orbital parameters as well as their Fourier transform.
The radiation-reaction corrections to $\bar a_r$ and $e_r$ lead to non-vanishing contributions to the radiated energy and angular momentum. 
We have checked that our results agree with known results on radiative losses computed along hyperboliclike orbits (see the pioneering work \cite{Blanchet:1989cu} as well as more recent ones \cite{Bini:2021gat,Bini:2022enm}). 

The knowledge of the radiation-reaction corrected orbit is  necessary for the calculation of the fractional 3.5PN corrections to the radiative losses as well as of the corresponding corrections to the radiative multipole moments entering the definition of the waveform.
This is left for future works.

\section*{Acknowledgments}

D.B. acknowledges
membership of the Italian Gruppo Nazionale per la Fisica
Matematica (GNFM) of the Istituto Nazionale di Alta
Matematica (INDAM). D.B. also acknowledges fruitful discussions with T. Damour and the hospitality and the
highly stimulating environment of the Institut des Hautes
Etudes Scientifiques at various stages during the development of the present project.  
S.R.A. ackowledges S. Capozziello and G. Lambiase for useful discussions. 
D.B. and S.R.A acknowledge L. Blanchet and G. Faye for valuable comments and suggestions.

\appendix

\begin{widetext}

\section{Total variation of the orbital parameters undergoing radiation reaction effects}
\label{jumps}

We list in Table \ref{tab:2} the total variation of the orbital parameters undergoing radiation reaction effects, say
\bea
[X^{\rm rr}]&=&[X^{\rm rr}]^{(0)}+\eta^2 [X^{\rm rr}]^{(2)}\nonumber\\
&=&\lim_{v\to +\infty} X^{\rm rr} -\lim_{v\to -\infty} X^{\rm rr}=\lim_{v\to +\infty} X^{\rm rr}\,,
\eea
since initial conditions at $-\infty$ were chosen so that $\lim_{v\to -\infty} X^{\rm rr}=0$ for all radiation-reaction modified orbital parameters.
We also recall the relation
\beq
{\rm arctan}(\alpha)\equiv{\rm arctan}\left(\sqrt{\frac{e_r+1}{e_r-1}}\right)
=\frac12 {\rm arccos}\left( -\frac{1}{e_r}\right)\,.
\eeq

% table 2
\begin{table*}  
\caption{\label{tab:2} Total variation of the orbital parameters undergoing radiation reaction effects: exact expressions in terms of $\bar a_r$ and $e_r$. 
}
\begin{ruledtabular}
\begin{tabular}{ll}
$[\delta^{rr}  \bar a_r]^{(0)}$ &$ \frac{\nu}{\bar a_r^{3/2}}\left[\frac{4 (673 e_r^2+602)}{ 45(e_r^2-1)^3}
+\frac{4 (37 e_r^4+292 e_r^2+96)}{15 (e_r^2-1)^{7/2}}{\rm arccos}\left(-\frac{1}{e_r}\right)\right]$\\
$[\delta^{rr}  \bar a_r]^{(2)}$&$ \frac{\nu}{\bar a_r^{5/2}}\left[\frac{ - [(49280\nu+707199) e_r^4+(671440 \nu+2300712) e_r^2+396480\nu+253284]}{3150(e_r^2-1)^4}\right.$\\
&$\left.-\frac{ [3823 e_r^6+(13776 \nu+98910) e_r^4+(48608\nu+110856) e_r^2+12096\nu+3824]}{ 210 (e_r^2-1)^{9/2}} {\rm arccos}\left(-\frac{1}{e_r}\right)\right]$\\
\hline
$[\delta^{rr}  e_r]^{(0)}$&$\frac{\nu}{\bar a_r^{5/2}}\left[ -\frac{2 (72 e_r^4+1069 e_r^2+134)}{45e_r(e_r^2-1)^2}
-\frac{2 e_r (121 e_r^2+304)}{ 15 (e_r^2-1)^{5/2}} {\rm arccos}\left(-\frac{1}{e_r}\right)\right]$\\
$[\delta^{rr}  e_r]^{(2)}$&$\frac{\nu}{\bar a_r^{7/2}}\left[  -\frac{[51840 e_r^6-(136640 \nu+1456529) e_r^4-(899920 \nu+1901522) e_r^2-80640\nu+45016]}{ 6300 e_r (e_r^2-1)^3}\right. $\\
&$\left.+\frac{ e_r [5281 e_r^4+(28560\nu+155100) e_r^2+45920\nu+57032]}{ 420 (e_r^2-1)^{7/2}} {\rm arccos}\left(-\frac{1}{e_r}\right) \right]$\\
\hline
$[\delta^{rr}  \bar n]^{(0)}$&$\frac{\nu}{\bar a_r^4}\left[ -\frac{2(673 e_r^2+602)}{15(e_r^2-1)^3}-\frac{2 (37 e_r^4+292 e_r^2+96)}{ 5(e_r^2-1)^{7/2}} {\rm arccos}\left(-\frac{1}{e_r}\right)\right]$\\
$[\delta^{rr}  \bar n]^{(2)}$&$\frac{\nu}{\bar a_r^5}\left[  \frac{ [(618940\nu-2118303) e_r^4+(1964620\nu+7349436) e_r^2+768040\nu+4552452]}{ 6300 (e_r^2-1)^4}\right.$\\ 
&$\left.+\frac{[(5180 \nu-35151) e_r^6+(77028\nu-24570) e_r^4+(118384\nu+579528) e_r^2+22848\nu+132432]}{ 420 (e_r^2-1)^{9/2}} {\rm arccos}\left(-\frac{1}{e_r}\right) \right]$\\
\hline
$[\delta^{rr}  K]^{(0)}$& $0$  \\
$[\delta^{rr}  K]^{(2)}$&$ \frac{48}{5}\frac{\nu}{\bar a_r^{7/2}}\left[ \frac{(2 e_r^2+13)}{(e_r^2-1)^3}
+\frac{(7 e_r^2+8)}{(e_r^2-1)^{7/2}} {\rm arccos}\left(-\frac{1}{e_r}\right) \right]$\\
\end{tabular}
\end{ruledtabular}
\end{table*}

For completeness, we list below the corresponding  expressions in terms of $(p_\infty, j)$ in a large-$j$ expansion limit.

\bea
[\delta^{rr}  \bar a_r] &=&\nu\left\{\left[
\left(\frac{2393}{420} -\frac{37}{5}\nu \right)\frac{\pi p_\infty^2}{ j^3}
+\left(\frac{166528}{1575} -\frac{6976}{45}\nu \right)\frac{p_\infty}{j^4}
+\left(-\frac{811}{5}\nu +\frac{17247}{140}\right)\frac{\pi}{j^5}\right.\right.\nonumber\\
&+&\left.\left(-\frac{48256}{45}\nu+\frac{316928}{315} \right)\frac{1}{ j^6 p_\infty}
+\left(\frac{7117}{12} -\frac{1505}{3}\nu\right)\frac{\pi}{ j^7p_\infty^2}
+\left(\frac{424448}{175}-\frac{123712}{75}\nu \right)\frac{1}{ j^8p_\infty^3}\right]\eta^2\nonumber\\
&+&\left.\frac{74\pi}{15j^3}
+\frac{3136}{ 45 p_\infty j^4}
+\frac{244\pi}{ 5 p_\infty^2 j^5}
+\frac{9344}{ 45 p_\infty^3 j^6}
+\frac{170\pi}{ 3 p_\infty^4 j^7}
+\frac{6208}{ 75 p_\infty^5 j^8}\right\} +O\left(\eta^4,\frac{1}{j^9}\right)
\eea
\bea
[\delta^{rr}  e_r] &=&
\nu \left\{\left[
\left(\frac{48}{35}  +\frac{8}{5}\nu\right)\frac{p_\infty^6}{j}
+\left(\frac{4019}{280}  +\frac{121}{15}\nu  \right)\frac{\pi p_\infty^5}{ j^2}
+\left(\frac{228616}{1575}  +\frac{5668}{45}\nu \right)\frac{p_\infty^4}{j^3}
+\left(\frac{534}{5}\nu  +\frac{18433}{336} \right)\frac{\pi p_\infty^3}{ j^4}\right.\right.\nonumber\\
&+&\left.\left(\frac{25727}{45}\nu -\frac{6582}{175} \right)\frac{p_\infty^2}{ j^5}
+\left(\frac{1679}{8}\nu -\frac{166469}{1344} \right)\frac{\pi p_\infty}{j^6}
+\left(\frac{25133}{50}\nu -\frac{33611}{63} \right)\frac{1}{j^7}\right]\eta^2 \nonumber\\
&-&\left.\frac{16}{5} \frac{p_\infty^4}{ j}
-\frac{121}{15} \frac{p_\infty^3\pi}{j^2}
-\frac{616}{9} \frac{p_\infty^2}{ j^3}
-\frac{971}{30} \frac{p_\infty \pi}{ j^4}
-\frac{1354}{15}\frac{1}{ j^5}
-\frac{1579}{120}\frac{\pi}{p_\infty j^6}
+\frac{259}{225}\frac{1}{ p_\infty^2 j^7}\right\} +O\left(\eta^4,\frac{1}{j^9}\right)\,.
\eea
\bea
[\delta^{rr}  \bar n]&=&\nu\left\{
\left[\left(-\frac{7573}{280}  +\frac{259}{15}\nu \right)\frac{\pi p_\infty^7}{j^3}
+\left(-\frac{73488}{175}  +\frac{14384}{45}\nu \right)\frac{p_\infty^6}{j^4}
+\left(-\frac{102981}{280} +\frac{3043}{10}\nu \right)\frac{\pi p_\infty^5}{j^5}\right.\right.\nonumber\\
&+&\left.\left(\frac{84064}{45}\nu -\frac{240224}{105} \right)\frac{p_\infty^4}{j^6}
+\left(\frac{2470}{3}\nu -\frac{8817}{8} \right)\frac{p_\infty^3\pi}{j^7}
+\left(-\frac{690992}{175} +\frac{193328}{75}\nu \right)\frac{p_\infty^2}{j^8}\right]
\eta^2\nonumber\\
&-& \left.\frac{37}{5} \frac{\pi p_\infty^5}{ j^3}
-\frac{1568}{15} \frac{p_\infty^4}{j^4}
-\frac{366}{5} \frac{p_\infty^3 \pi}{ j^5}
-\frac{4672}{15} \frac{p_\infty^2}{j^6}
-85 \frac{p_\infty \pi}{j^7}
-\frac{3104}{25} \frac{1}{j^8}\right\} +O\left(\eta^4,\frac{1}{j^9}\right)\,.
\eea
\bea
[\delta^{rr}  K]&=&
\nu\left(\frac{96}{5} \frac{p_\infty^3}{j^4}+\frac{168}{5}\pi \frac{p_\infty^2}{j^5}+\frac{1056}{5} \frac{p_\infty}{j^6}+72\frac{\pi}{j^7}+\frac{608}{5  p_\infty j^8}\right)\eta^2 +O\left(\eta^4,\frac{1}{j^9}\right)
\eea
Inserting these total changes into the relations with the energy and angular momentum losses we have checked previous results, see e.g., Ref. \cite{Bini:2021gat}.
\end{widetext}

\end{document}